\pdfoutput=1
\documentclass[11pt,twocolumn]{article}

\usepackage{times}
\usepackage[
    left=1.75cm,
    right=1.75cm,
    top=3.00cm,
    bottom=3.00cm
]{geometry}

\usepackage{mathptmx}
\usepackage{xcolor}
\usepackage{framed}
\definecolor{shadecolor}{HTML}{F0F0F0}
\usepackage[inline]{enumitem}
\usepackage[abbreviations]{foreign}  
\usepackage{graphicx} 
\usepackage[font={small,stretch=1.0},skip=2pt,belowskip=4pt,labelfont=bf]{caption} 
\usepackage[utf8]{inputenc}
\usepackage{multirow,tabularx} 
\usepackage{epsfig}
\usepackage{booktabs} 
\usepackage[T1]{fontenc}
\usepackage{times}
\usepackage{pifont}
\usepackage{algorithm}
\usepackage{algpseudocode}
\usepackage{listings}
\usepackage{xspace}
\usepackage{ifthen}
\usepackage{amsmath,amssymb,bm}
\usepackage{array}

\usepackage[acronym]{glossaries}

\usepackage[scaled=0.85]{beramono}

\usepackage[normalem]{ulem}
\usepackage{setspace}
\usepackage{hyperref} 

\newcommand\YAMLcolonstyle{\color{red!70!black}\small\small\ttfamily}
\newcommand\YAMLkeystyle{\color{black}\small\small\ttfamily\bfseries}
\newcommand\YAMLvaluestyle{\color{black}\small\small\ttfamily}
\newcommand\YAMLcommentstyle{\color{gray!80!black}\scriptsize\scriptsize\ttfamily\itshape}
\makeatletter
\newcommand\language@yaml{yaml}
\expandafter\expandafter\expandafter\lstdefinelanguage
\expandafter{\language@yaml}
{
	keywords={true,false,null,y,n},
	keywordstyle=\color{darkgray}\bfseries,
	basicstyle=\linespread{0.6}\small\ttfamily\color{black},  
	sensitive=false,
	comment=[l]{\#},
	morecomment=[s]{/*}{*/},
	commentstyle=\YAMLcommentstyle,
	stringstyle=\YAMLvaluestyle,
	moredelim=[l][\color{orange}]{\&},
	moredelim=[l][\color{magenta}]{*},
	moredelim=**[il][\YAMLcolonstyle{:}\YAMLvaluestyle]{},  
	morestring=[b]',
	morestring=[b]",
	literate = {---}{{\ProcessThreeDashes}}3
	{>}{{\textcolor{red!70!black}\textgreater}}1
	{|}{{\textcolor{red!70!black}\textbar}}1
	{-}{{\mdseries-}}1,
}

\lst@AddToHook{EveryLine}{\ifx\lst@language\language@yaml\YAMLkeystyle\fi}
\makeatother

\lstdefinelanguage{Ini}
{
	basicstyle=\fontsize{10}{10}\ttfamily,
	morecomment=[s][\color{blue}]{[}{]},
	morecomment=[l]{\#},
	morecomment=[l]{;},
	commentstyle=\color{gray},
	morekeywords={},
	otherkeywords={=},
	keywordstyle={\color{gray}}
}

\newacronym{tbe}{TBE}{trusted boot enclave}
\newacronym{sgx}{SGX}{Intel software guard extensions}
\newacronym{txt}{TXT}{Intel trusted execution technology}
\newacronym{tee}{TEE}{trusted execution environment}
\newacronym{kms}{KMS}{key management system}
\newacronym{epc}{EPC}{enclave page cache}
\newacronym{tls}{TLS}{transport layer security}
\newacronym{os}{OS}{operating system}
\newacronym{drtm}{DRTM}{dynamic root of trust for measurements}
\newacronym{srtm}{SRTM}{static root of trust for measurements}
\newacronym{crtm}{CRTM}{core root of trust for measurements}
\newacronym{rom}{ROM}{read-only memory}
\newacronym{ram}{RAM}{random-access memory}
\newacronym{cpu}{CPU}{central processing unit}
\newacronym{nvram}{NVRAM}{non-volatile random-access memory}
\newacronym{dram}{DRAM}{dynamic random-access memory}
\newacronym{ek}{EK}{endorsement key}
\newacronym{aik}{AIK}{attestation key}
\newacronym{ca}{CA}{certificate authority}
\newacronym{tpm}{TPM}{trusted platform module}
\newacronym{dtpm}{dTPM}{discrete TPM chip}
\newacronym{ftpm}{fTPM}{firmware TPM}
\newacronym{vtpm}{vTPM}{virtual TPM}
\newacronym{itpm}{iTPM}{integrated TPM}
\newacronym{pch}{PCH}{platform controller hub}
\newacronym{tcg}{TCG}{Trusted Computer Group}
\newacronym{pcr}{PCR}{platform configuration register}
\newacronym{pcrd}{dynamic PCR}{dynamic PCR}
\newacronym{pcrs}{static PCR}{static PCR}
\newacronym{ptt}{PTT}{Intel platform trusted technology}
\newacronym{uefi}{UEFI}{unified extensible firmware interface}
\newacronym{bios}{BIOS}{basic input/output system}
\newacronym{pxe}{PXE}{preboot execution environment}
\newacronym{svm}{SVM}{Secure Virtual Machine}
\newacronym{ima}{IMA}{integrity measurement architecture}
\newacronym{vpn}{VPN}{virtual private network}
\newacronym{daa}{DAA}{direct anonymous attestation}
\newacronym{loc}{LOC}{lines of code}
\newacronym{sloc}{SLOC}{source lines of code}
\newacronym{mee}{MEE}{memory encryption engine}
\newacronym{ias}{IAS}{Intel attestation service}
\newacronym{mrenclave}{MRENCLAVE}{enclave hash measurement}
\newacronym{acs}{IBM ACS}{IBM TPM attestation client-server}
\newacronym{cit}{Intel CIT}{Intel open cloud integrity technology}
\newacronym{initramfs}{initramfs}{initramfs}
\newacronym{vm}{VM}{virtual machine}
\newacronym{iaas}{IaaS}{Infrastructure-as-a-Service}
\newacronym{maas}{MaaS}{Metal-as-a-Service}
\newacronym{lpc}{LPC}{low pin count}
\newacronym{me}{CSME}{Intel converged security and manageability engine}
\newacronym{toctou}{TOCTOU}{time of check to time of use}
\newacronym{itl}{ITL}{Invisible Things Lab}
\newacronym{smm}{SMM}{system management mode}
\newacronym{dma}{DMA}{direct memory access}
\newacronym{tcb}{TCB}{trusted computing base}
\newacronym{bmc}{BMC}{baseboard management controller}
\newacronym{ipmi}{IPMI}{intelligent platform management interface}
\newacronym{nic}{NIC}{network interface card}
\newacronym{ssh}{SSH}{secure shell}
\newacronym{hsm}{HSM}{hardware security module}
\newacronym{vmm}{VMM}{virtual machine monitor}
\newacronym{kvm}{KVM}{kernel-based virtual machine}
\newacronym{qemu}{QEMU}{quick emulator}
\newacronym{mktme}{MKTME}{Intel multi-key total memory encryption}
\newacronym{tdx}{TDX}{Intel trust domain extensions}
\newacronym{tme}{TME}{Total Memory Encryption}
\newacronym{isp}{ISP}{internet service provider}
\newacronym{ssl}{SSL}{secure sockets layer}
\newacronym{ecc}{ECC}{elliptic-curve cryptography}
\newacronym{rsa}{RSA}{Rivest Shamir Adleman}
\newacronym{vmbr}{VMBR}{virtual-machine based rootkit}
\newacronym{aes}{AES}{advanced encryption standard}
\newacronym{sev}{SEV}{AMD secure encrypted virtualization}
\newacronym{ip}{IP}{internet protocol}
\newacronym{dns}{DNS}{domain name system}
\newacronym{mitm}{MitM}{man-in-the-middle}
\newacronym{arp}{ARP}{address resolution protocol}
\newacronym{tcp}{TCP}{transmission control protocol}
\newacronym{mc}{MC}{monotonic counter}
\newacronym{mcs}{MCS}{monotonic counter service}
\newacronym{rest}{REST}{representational state transfer}
\newacronym{api}{API}{application programming interface}
\newacronym{cve}{CVE}{common vulnerabilities and exposures}
\newacronym{sriov}{SR-IOV}{single root input/output virtualization}
\newacronym{vtd}{VT-d}{Intel virtualization technology for directed I/O}
\newacronym{ecdsa}{ECDSA}{elliptic curve digital signature algorithm}
\newacronym{pal}{PAL}{piece of application logic}
\newacronym{iommu}{IOMMU}{input-output memory management unit}
\glsdisablehyper
\makeglossaries

\newcommand{\sysnospace}{{\rm\textsc{Weles}}}
\newcommand{\sys}{\sysnospace\xspace}

\newcommand{\imalog}{IMA log\xspace}

\newcommand{\captionvspacesize}{0mm}
\newcommand{\captiontablevspacesize}{-2mm}
\newcommand{\myparagraph}[1]{\vspace{1mm} \smallskip \noindent{\bf {#1}}}
\newcommand{\specialcell}[2][c]{\begin{tabular}[#1]{@{}c@{}}#2\end{tabular}}

\begin{document}

\title{\sysnospace: Policy-driven Runtime Integrity Enforcement of Virtual Machines}

\author{
        Wojciech Ozga\\
        \textit{TU Dresden, Germany} \\
        \textit{IBM Research Europe}
     \and
        Do Le Quoc\\
        \textit{TU Dresden, Germany}
     \and
        Christof Fetzer\\
        \textit{TU Dresden, Germany}
}

\maketitle

\begin{abstract}
	
Trust is of paramount concern for tenants to deploy their security-sensitive services in the cloud. 
The integrity of \glspl{vm} in which these services are deployed needs to be ensured even in the presence of powerful adversaries with administrative access to the cloud. 
Traditional approaches for solving this challenge leverage trusted computing techniques, \eg, vTPM, or hardware CPU extensions, \eg, AMD SEV.
But, they are vulnerable to powerful adversaries, or they provide only load time (not runtime) integrity measurements of VMs.

We propose \sys, a protocol allowing tenants to establish and maintain trust in VM \emph{runtime integrity} of software and its configuration.
\sys is transparent to the VM configuration and setup. 
It performs an implicit attestation of VMs during a secure login and binds the VM integrity state with the secure connection. 
Our prototype's evaluation shows that \sys is practical and incurs low performance overhead ($\le 6\%$).

\glsresetall
\end{abstract}
\section{Introduction}
\label{sec:introduction}

Cloud computing paradigm shifts the responsibility of the computing resources management from application owners to cloud providers, allowing application owners (tenants) to focus on their business use cases instead of on hardware management and administration. However, trust is of paramount concern for tenants operating security-sensitive systems because software managing computing resources and its configuration and administration remains out of their control.
Tenants have to trust that the cloud provider, its employees, and the infrastructure protect the tenant's intellectual property as well as the confidentiality and the integrity of the tenant's data.
A malicious employee~\cite{privacy_google_fired}, or an adversary who gets into possession of employee credentials~\cite{ibm_xforce_threat_intelligence_index_2020, nsa_admins}, might leverage administrator privileges to read the confidential data by introspecting \gls{vm} memory~\cite{vm_introspecting_2015}, to tamper with computation by subverting the hypervisor~\cite{king_subvirt:_2006}, or to redirect the tenant to an arbitrary \gls{vm} under her control by altering a network configuration \cite{zheng_application-based_2009}. 
We tackle the problem of how to establish trust in a VM executed in the cloud. Specifically, we focus on the integrity of legacy systems executed in a VM.

The existing attestation protocols focus on leveraging trusted hardware to report measurements of the execution environment. In trusted computing~\cite{trusted_computing_2009}, the trusted platform module attestation~\cite{tpm_2_0_spec} and \gls{ima}~\cite{ima_design_2004} provide a means to enforce and monitor integrity of the software that has been executed since the platform bootstrap~\cite{tcg_ima_spec}. The \gls{vtpm}~\cite{perez_vtpm:_2006} design extends this concept by introducing a software-based \gls{tpm} that, together with the hardware \gls{tpm}, provides integrity measurements of the entire software stack --- from the firmware, the hypervisor, up to the \gls{vm}. However, this technique cannot be applied to the cloud because an adversary can tamper with the communication between the \gls{vtpm} and the VM. 
For example, by reconfiguring the network, she can mount a man-in-the-middle attack to perform a TPM reset attack~\cite{kauer_oslo_2007}, compromising the vTPM security guarantees.

A complementary technology to trusted computing, \gls{tee}~\cite{tee_whitepaper}, uses hardware extensions to exclude the administrator and privileged software, \ie, operating system, hypervisor, from the trusted computing base. The \gls{sgx}~\cite{costan2016intel} comes with an attestation protocol that permits remotely verifying application's integrity and the genuineness of the underlying hardware. However, it is available only to applications executing inside an \gls{sgx} enclave. Legacy applications executed inside an enclave suffer from performance limitations due to a small amount of protected memory \cite{arnautov_scone:_2016}. The SGX adoption in the virtualized environment is further limited because the protected memory is shared among all tenants. 

Alternative technologies isolating VMs from the untrusted hypervisor, \eg, AMD SEV~\cite{amd_sme_whitepaper, amd_sev_es_whitepaper} or IBM PEF~\cite{hunt2021confidential}, do not have memory limitations. They support running the entire operating system in isolation from the hypervisor while incurring minimal performance overhead~\cite{gottel_security_2018}. However, their attestation protocol only provides information about the \gls{vm} integrity \emph{at the \gls{vm} initialization time}. It is not sufficient because the loaded operating system might get compromised later--at \emph{runtime}--with operating system vulnerabilities or misconfiguration~\cite{configuration_error_analysis}. Thus, to verify the runtime (post-initialization) integrity of the guest operating system, one would still need to rely on the \gls{vtpm} design. But, as already mentioned, it is not enough in the cloud environment.

Importantly, security models of these hardware technologies isolating VM from the hypervisor assume threats caused by running tenants' OSes in a shared execution environment, \ie, attacks performed by rogue operators, compromised hypervisor, or malicious co-tenants. These technologies do not address the fact that a typical tenant's OS is a complex mixture of software and configuration with a large vector attack. \emph{I.e.}, the protected application is not, like in the \gls{sgx}, a single process, but the kernel, userspace services, and applications, which might be compromised while running inside the TEE and thus exposes tenant's computation and data to threats. These technologies assume it is the tenant's responsibility to protect the OS, but they lack primitives to enable runtime integrity verification and enforcement of guest OSes. This work proposes means to enable such primitives, which are neither provided by the technologies mentioned above nor by the existing cloud offerings. 

We present \sysnospace, a \gls{vm} remote attestation protocol that provides integrity guarantees to legacy systems executed in the cloud.
\sys has noteworthy advantages. 
First, it supports legacy systems with zero-code changes by running them inside \glspl{vm} on the integrity-enforced execution environment. To do so, it leverages trusted computing to enforce and attest to the hypervisor's and \gls{vm}'s integrity.
Second, \sys limits the system administrator activities in the host OS using integrity-enforcement mechanisms, while relying on the \gls{tee} to protect its own integrity from tampering.
Third, it supports tenants connecting from machines not equipped with trusted hardware. 
Specifically, \sys integrates with the \gls{ssh} protocol~\cite{ssh_standard}. Login to the \gls{vm} implicitly performs an attestation of the \gls{vm}. 

Our contributions are as follows:
\setlist{nolistsep}
\begin{itemize}[noitemsep, leftmargin=4mm]
    \item We designed a protocol, \sys, attesting to the VM's runtime integrity (\S\ref{sec:overview}).
    \item We implemented the \sys prototype using state-of-the-art technologies commonly used in the cloud (\S\ref{sec:implementation}).
    \item We evaluated it on real-world applications (\S\ref{sec:evaluation}).
\end{itemize}

\section{Threat model}
\label{sec:threat_model}

We require that the \emph{cloud node} is built from the software which source code is certified by a trusted third party~\cite{nist_hashes} or can be reviewed by tenants, \eg, open-source software~\cite{firecracker} or proprietary software accessible under non-disclosure agreement. 
Specifically, such software is typically considered safe and trusted when
\begin{enumerate*}[label=(\roman*)]
    \item it originates from trusted places like the official Linux git repository;
    \item it passes security analysis like fuzzing~\cite{fuzzing}; 
    \item it is implemented using memory safe languages, like Rust; 
    \item it has been formally proven, like seL4~\cite{klein_sel4:_2009} or EverCrypt~\cite{protzenko2020evercrypt}; 
    \item it was compiled with memory corruption mitigations, \eg, position-independent executables with stack-smashing protection.
\end{enumerate*}

Our goal is to provide tenants with an \emph{runtime integrity attestation protocol} that ensures that the cloud node (\ie, host OS, hypervisor) and the VM (guest OS, tenant's legacy application) run only expected software in the expected configuration. 
We distinguish between an internal and an external adversary, both without capabilities of mounting physical and hardware attacks (\eg, \cite{winter_hijackers_2013}). This is a reasonable assumption since cloud providers control and limit physical access to their data centers.

An \emph{internal adversary}, such as a malicious administrator or an adversary who successfully extracted administrators credentials~\cite{ibm_xforce_threat_intelligence_index_2020, nsa_admins}, aims to tamper with a hypervisor configuration or with a VM deployment to compromise the integrity of the tenant's legacy application. She has remote administrative access to the host machine that allows her to configure, install, and execute software. The internal adversary controls the network that will allow her to insert, alter, and drop network packages. 

An \emph{external adversary} resides outside the cloud. Her goal is to compromise the security-sensitive application's integrity. She can exploit a guest OS misconfiguration or use social engineering to connect to the tenant's VM remotely. Then, she runs dedicated software, \eg, software debugger or custom kernel, to modify the legacy application's behavior.

We consider the \gls{tpm}, the CPU, and their hardware features trusted.
We rely on the soundness of cryptographic primitives used by software and hardware components.
We treat software-based side-channel attacks (\eg, \cite{Kocher2018spectre}) as orthogonal to this work because of (i) the counter-measures existence (\eg, \cite{varys_2018}) whose presence is verifiable as part of the \sys protocol, (ii) the possibility of provisioning a dedicated (not shared) machine in the cloud.

\section{Background and Problem Statement}
\label{sec:background}

\myparagraph{Load-time integrity enforcement.}
A cloud node is a computer where multiple tenants run their \glspl{vm} in parallel on top of the same computing resources. 
VMs are managed by a hypervisor, a privileged layer of software providing access to physical resources and isolating \glspl{vm} from each other. 
Since the \gls{vm}'s security depends on the hypervisor, it is essential to ensure that the correct hypervisor controls the \gls{vm}.

The trusted computing~\cite{trusted_computing_2009} provides hardware and software technologies to verify the hypervisor's integrity. 
In particular, the \gls{drtm}~\cite{drtm_tcg} is a mechanism available in modern CPUs that establishes a trusted environment in which a hypervisor is securely measured and loaded.
The hypervisor's integrity measurements are stored inside the hardware \gls{tpm} chip in dedicated memory regions called \glspl{pcr}~\cite{tpm_2_0_spec}. 
\glspl{pcr} are tamper-resistant. 
They \emph{cannot be written directly} but only extended with a new value using a cryptographic hash function: \emph{PCR\_extend = hash(PCR\_old\_value $\vert\vert$ data\_to\_extend)}.

The TPM attestation protocol~\cite{tcg_tpm_attestation} defines how to read a report certifying the PCRs values.
The report is signed using a cryptographic key derived from the endorsement key, which is an asymmetric key embedded in the TPM chip at the manufacturing time.
The TPM also stores a certificate, signed by a manufacturer, containing the endorsement key's public part.
Consequently, it is possible to check that a genuine TPM chip produced the report because the report's signature is verifiable using the public key read from the certificate.

\myparagraph{Runtime integrity enforcement.}
The administrator has privileged access to the machine with complete control over the network configuration, with permissions to install, start, and stop applications. 
These privileges permit him to trick the \gls{drtm} attestation process because the hypervisor's integrity is measured \emph{just once} when the hypervisor is loaded to the memory.
The TPM report certifies this state until the next \gls{drtm} launch, \ie, the next computer boot.
Hence, after the hypervisor has been measured, an adversary can compromise it by installing an arbitrary hypervisor~\cite{bluepill:2006} or downgrading it to a vulnerable version without being detected.

\Acrfull{ima}~\cite{tcg_ima_spec, ima_design_2004, ima_appraisal} allows for mitigation of the threat mentioned above. 
Being part of the measured kernel, IMA implements an \emph{integrity-enforcement mechanism}~\cite{ima_appraisal}, allowing for loading only digitally signed software and configuration.
Consequently, signing only software required to manage VMs allows for limiting activities carried out by an administrator on the host machine.
A load of a legitimate kernel with enabled IMA and \glsdesc{iommu} is ensured by \gls{drtm}, and it is attestable via the \gls{tpm} attestation protocol.

\myparagraph{Integrity auditing.}
IMA provides information on what software has been installed or launched since the kernel loading, what is the system configuration, and whether the system configuration has changed. 
\gls{ima} provides a tamper-proof history of collected measurements in a dedicated file called \imalog.
The tamper-proof property is maintained because IMA extends to the PCR the measurements of every executable, configuration file, script, and dynamic library before such file or program is read or executed. 
Consequently, an adversary who tampers with the \imalog cannot hide her malicious behavior because she cannot modify the PCR value. 
She cannot overwrite the PCR and she cannot reset it to the initial value without rebooting the platform.

\label{sec:vtpm_vulnerabilities}
\label{sec:vtpm_attacks}
\begin{figure}[btp!]
    \centering
    \includegraphics[width=0.48\textwidth]{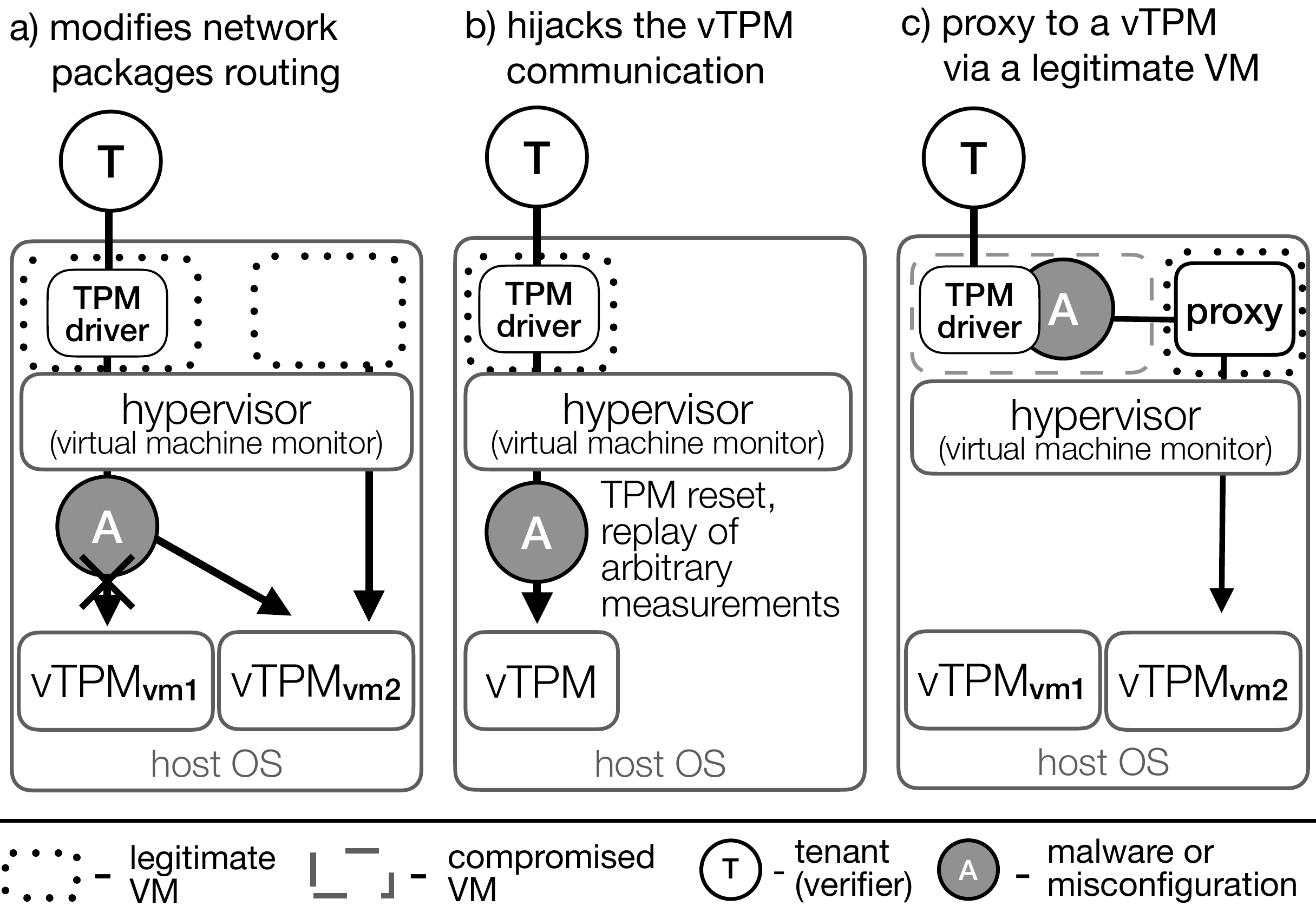}
    \vspace{\captionvspacesize}
    \caption{     
        An adversary with root access to the hypervisor can violate the security guarantees promised by the vTPM~\cite{perez_vtpm:_2006} design.
    }
    \label{fig:vtpm_attacks}
\end{figure}

\myparagraph{Problems with virtualized TPMs.}
The \gls{tpm} chip cannot be effectively shared with many \glspl{vm} due to a limited amount of \glspl{pcr}.
The \gls{vtpm}~\cite{perez_vtpm:_2006} design addresses this problem by running multiple software-based \glspl{tpm} exposed to \glspl{vm} by the hypervisor. 
This design requires verifying the hypervisor's integrity before establishing trust with a software-based \gls{tpm}.
We argue that verifying the hypervisor's integrity alone is not enough because an administrator can break the software-based \gls{tpm} security guarantees by mounting a \gls{mitm} attack using the legitimate software, as we describe next. Consequently, the vTPM cannot be used directly to provide runtime integrity of VMs.

In the \gls{vtpm} design, the hypervisor prepends a 4-byte \gls{vtpm} identifier that allows routing the communication to the correct \gls{vtpm} instance. However, the link between the \gls{vtpm} and the VM is unprotected~\cite{cucurull_virtual_2014}, and it is routed through an untrusted network.
Consequently, an adversary can mount a masquerading attack to redirect the VM communication to an arbitrary \gls{vtpm} (\autoref{fig:vtpm_attacks}a) by replacing the \gls{vtpm} identifier inside the network package. 
To mitigate the attack, we propose to use the \gls{tls} protocol~\cite{tls_1_2} to protect the communication's integrity.

Although the \gls{tls} helps protect the communication's integrity, lack of authentication between the \gls{vtpm} and the hypervisor still enables an adversary to fully control the communication by placing a proxy in front of the vTPM.
In more detail, an adversary can configure the hypervisor in a way it communicates with vTPM via an intermediary software, which intercepts the communication (\autoref{fig:vtpm_attacks}b). She can then drop arbitrary measurements or perform the \gls{tpm} reset attack~\cite{kauer_oslo_2007}, thus compromising the vTPM security guarantees.

To mitigate the attack, the vTPM must ensure the remote peer's integrity (is it the correct hypervisor?) and its locality (is the hypervisor running on the same platform?). Although the TEE local attestation gives information about software integrity and locality, we cannot use it here because the hypervisor cannot run inside the TEE. However, suppose we find a way to satisfy the locality condition. In that case, we can leverage integrity measurements (IMA) to verify the hypervisor's integrity because among trusted software running on the platform there can be only one that connects to the vTPM---the hypervisor. 
To satisfy the locality condition, we make the following observation: Only software running on the same platform has direct access to the same hardware \gls{tpm}. 
We propose to share a secret between the vTPM and the hypervisor using the hardware \gls{tpm} (\S\ref{sec:impl:qemu}).
The vTPM then authenticates the hypervisor by verifying that the hypervisor presents the secret in the pre-shared key \gls{tls} authentication.

Finally, an adversary who compromises the guest OS can mount the cuckoo attack~\cite{parno_bootstrapping} to impersonate the legitimate VM.
In more detail, an adversary can modify the TPM driver inside a guest OS to redirect the TPM communication to a remote TPM (\autoref{fig:vtpm_attacks}c).
A verifier running inside a compromised \gls{vm} cannot recognize if he communicates with the \gls{vtpm} attached to his \gls{vm} or with a remote \gls{vtpm} attached to another \gls{vm}. 
The verifier is helpless because he cannot establish a secure channel to the vTPM that would guarantee communication with the local \gls{vtpm}. 
To mitigate the attack, we propose leveraging the TEE attestation protocol to establish a secure communication channel between the verifier and the vTPM and to use it to exchange a secret allowing the verifier to identify the vTPM instance uniquely (\S\ref{sec:establish_trust}).

\section{\sys design}
\label{sec:overview}
Our objective is to provide an architecture that: 
\setlist{nolistsep}
\begin{enumerate*}[noitemsep, leftmargin=4mm]
\item protects legacy applications running inside a \gls{vm} from threats defined in \S\ref{sec:threat_model},
\item requires zero-code changes to legacy applications and the \gls{vm} setup,
\item permits tenants to remotely attest to the platform runtime integrity without possessing any vendor-specific hardware.
\end{enumerate*}

\subsection{High-level Overview}
\begin{figure}[tbp!]
\centering
\includegraphics[width=0.48\textwidth]{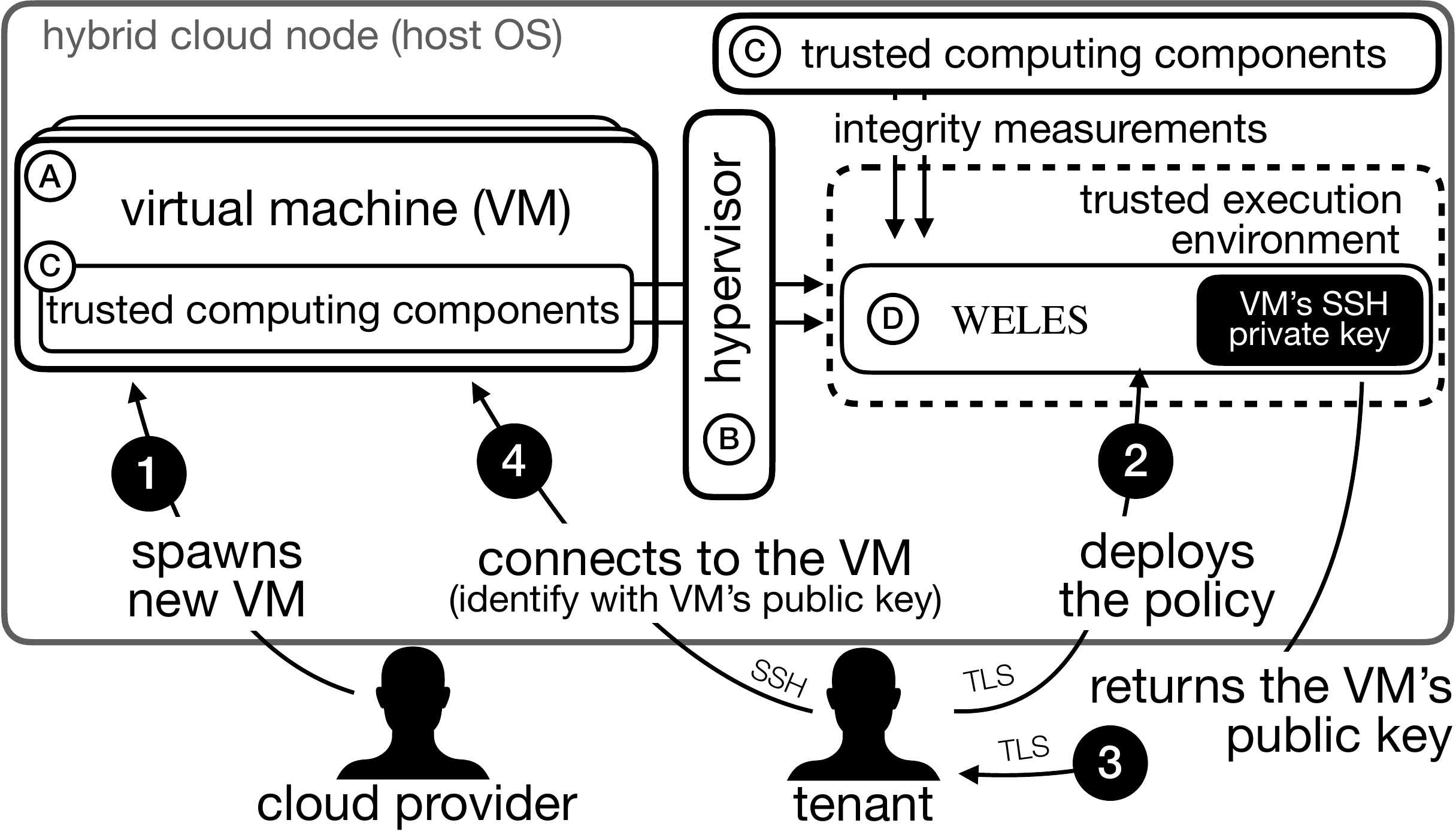}
\vspace{\captionvspacesize}
\caption{
    The high-level overview of \sys. 
    The VM's SSH key is bound to the VM's integrity state defined in the policy.
}
\label{fig:architecture}
\end{figure}
\autoref{fig:architecture} shows an overview of the cloud node running \sys.
It consists of the following four components: 
(A) the \gls{vm},
(B) the hypervisor managing the \gls{vm}, providing it with access to physical resources and isolating from other \glspl{vm},
(C) trusted computing components enabling hypervisor's runtime integrity enforcement and attestation,
(D) \sys, software executed inside \gls{tee} that allows tenants to attest and enforce the \glspl{vm}' integrity.

The configuration, the execution, and the operation of the above components are subject to attestation. First, the cloud operator bootstraps the cloud node and starts \sys. At the tenant's request, the cloud provider spawns a \gls{vm} (\raisebox{-1pt}{\ding{202}}). Next, the tenant establishes trust with \sys (\S\ref{sec:establish_trust}), which becomes the first trusted component on a remote platform.
The tenant requests \sys to check if the hypervisor conforms to the policy (\raisebox{-1pt}{\ding{203}}), which contains tenant-specific trust constraints, such as integrity measurements (\S\ref{sec:policy}). 
\sys uses IMA and TPM to verify that the platform's runtime integrity conforms to the policy and then generates a VM's public/private key pair. The public key is returned to the tenant (\raisebox{-1pt}{\ding{204}}). \sys protects access to the private key, \ie, it permits the \gls{vm} to use the private key only if the platform matches the state defined inside the policy. Finally, the tenant establishes trust with the \gls{vm} during the \gls{ssh}-handshake. He verifues that the VM can use the private key corresponding to the previously obtained public key (\raisebox{-1pt}{\ding{205}}).
The tenant authenticates himself in a standard way, using his own SSH private key. His SSH public key is embedeed inside the VM's image or provisioned during the VM deployment.

\begin{figure}[tbp!]
\centering
\includegraphics[width=0.48\textwidth]{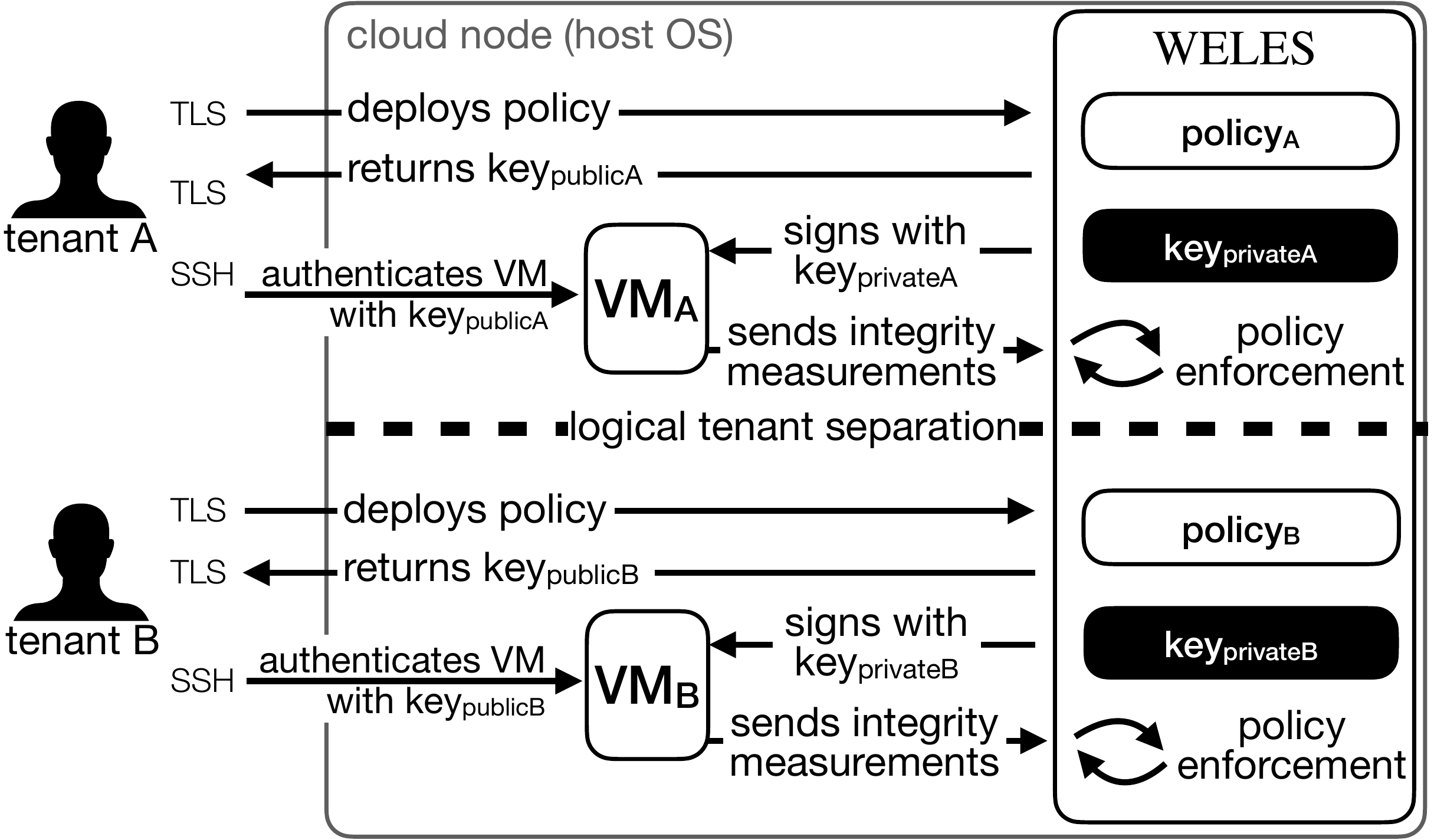}
\vspace{\captionvspacesize}
\caption{
    Multiple tenants interacting with \sys concurrently. 
    \sys generates a dedicated SSH key for each deployed security policy and allows using it only if the VM's integrity conforms to the security policy.
}
\label{fig:multiple_vms}
\end{figure}
\subsection{Tenant Isolation and Security Policy}
\label{sec:policy}
Multiple applications with different security requirements might coexist on the same physical machine. \sys allows ensuring that applications run in isolation from each other and match their security requirements. \autoref{fig:multiple_vms} shows how \sys assigns each VM a pair of a public and private key. The keys are bound with the application's policy and the VM's integrity. Each tenant uses the public key to verify that he interacts with his VM controlled by the integrity-enforced hypervisor.

\autoref{lst:policy} shows an example of a security policy. 
The policy is a text file containing a whitelist of the hardware TPM manufacturer's certificate chain (line \ref{policy_tpm_cert}), DRTM integrity measurements of the host kernel (lines \ref{policy_dynamic_pcrs_start}-\ref{policy_dynamic_pcrs_end}), integrity measurements of the guest kernel (line \ref{guest_pcrs}), and legal runtime integrity measurements of the guest OS (lines \ref{policy_vm_ima_start}-\ref{policy_vm_ima_end}, \ref{policy_vm_ima_cert}). 
The certificate chain is used to establish trust in the underlying hardware \gls{tpm}. 
\sys compares DRTM integrity measurements with PCR values certified by the TPM to ensure the correct hypervisor with enabled integrity-enforced mechanism was loaded. 
\sys uses runtime integrity measurements to verify that only expected files and software have been loaded to the guest OS memory.
A dedicated certificate (line \ref{policy_vm_ima_cert}) makes the integrity-enforcement practical because it permits more files to be loaded to the memory without redeploying the policy. Specifically, it is enough to sign software allowed to execute with the corresponding private key to let the software pass through the integrity-enforcement mechanism. Additional certificates (lines \ref{policy_ima_cert2}, \ref{policy_vm_ima_cert2}) allow for OS updates~\cite{tsr_2020}.

\subsection{Platform Bootstrap}
The cloud provider is responsible for the proper machine initialization. She must turn on support for hardware technologies (\ie, \gls{tpm}, \gls{drtm}, \gls{tee}), launch the hypervisor, and start \sys. Tenants detect if the platform was correctly initialized when they establish trust in the platform (\S\ref{sec:establish_trust}).

First, \sys establishes a connection with the hardware TPM using the \gls{tpm} attestation; it reads the \gls{tpm} certificate and generates an attestation key following the \emph{activation of credential} procedure (\cite{arthur_practical_2015} p. 109-111). 
\sys ensures it communicates with the local TPM using \cite{tbe_2019}, but other approaches might be used as well~\cite{mccune_catching_2011, parno_bootstrapping}.
Eventually, \sys reads the TPM quote, which certifies the DRTM launch and the measurements of the hypervisor's integrity.

\lstdefinestyle{interfaces}{
  float=tbp!,
  floatplacement=tbp
}
\begin{lstlisting}[caption=Example of the \sys's security policy,label={lst:policy},language=yaml,breaklines=true,breakatwhitespace=true,style=interfaces,escapechar=^]
host: ^\label{policy_host_start}^
    tpm: |- 
        -----BEGIN CERTIFICATE-----
        # Manufacturer CA certificate ^\label{policy_tpm_cert}^
        -----END CERTIFICATE-----
    drtm:   # measurements provided by the DRTM ^\label{policy_dynamic_pcrs_start}^
        - {id: 17, sha256: f9ad0...cb}
        - {id: 18, sha256: c2c1a...c1}
        - {id: 19, sha256: a18e7...00}      ^\label{policy_dynamic_pcrs_end}^
    certificate: |-
            -----BEGIN CERTIFICATE-----
            # software update certificate ^\label{policy_ima_cert2}^
            -----END CERTIFICATE----- 
guest: ^\label{policy_guest_start}^
    enforcement: true   ^\label{enforcement}^
    pcrs:   # boot measurements (e.g., secure boot)  ^\label{guest_pcrs}^
        - {id: 0, sha256: a1a1f...dd}
    measurements:   # legal integrity measurement digests ^\label{policy_vm_ima_start}^
        - "e0a11...2a" # SHA digest over a startup script
        - "3a10b...bb" # SHA digest over a library ^\label{policy_vm_ima_end}^
    certificate:
        - |-
            -----BEGIN CERTIFICATE-----
            # certificate of the signer, e.g., OS updates  ^\label{policy_vm_ima_cert}^
            -----END CERTIFICATE----- 
            -----BEGIN CERTIFICATE-----
            # software update certificate ^\label{policy_vm_ima_cert2}^
            -----END CERTIFICATE-----   ^\label{policy_guest_end}^            
\end{lstlisting}

\subsection{VM Launch}
\label{sec:impl:qemu}
The cloud provider requests the hypervisor to spawn a new VM. The hypervisor allocates the required resources and starts the VM providing it with TPM access. At the end of the process, the cloud provider shares the connection details with the tenant, allowing the tenant to connect to the \gls{vm}.

\sys emulates multiple \glspl{tpm} inside the \gls{tee} because many \glspl{vm} cannot share a single hardware \gls{tpm} \cite{perez_vtpm:_2006}. When requested by the hypervisor, \sys spawns a new \gls{tpm} instance accessible on a unique TCP port. The hypervisor connects to the emulated \gls{tpm} and exposes it to the \gls{vm} as a standard character device. We further use the term \emph{emulated TPM} to describe a TEE-based TPM running inside the hypervisor and distinguish it from the software-based TPM proposed by the vTPM design.

The communication between the hypervisor and the emulated \gls{tpm} is susceptible to \gls{mitm} attacks. Unlike \sys, the hypervisor does not execute inside the \gls{tee}, preventing \sys from using the \gls{tee} attestation to verify the hypervisor identity. However, \sys confirms the hypervisor identity by requesting it to present a secret when establishing a connection. \sys generates a secret inside the \gls{tee} and seals it to the hardware \gls{tpm} via an encrypted channel (\cite{tpm_2_0_spec_architecture}~\S19.6.7). Only software running on the same OS as \sys can unseal the secret. Thus, it is enough to check if only trusted software executes on the platform to verify that it is the legitimate hypervisor who presents the secret.

\begin{figure}[btp!]
\centering
\includegraphics[width=0.48\textwidth]{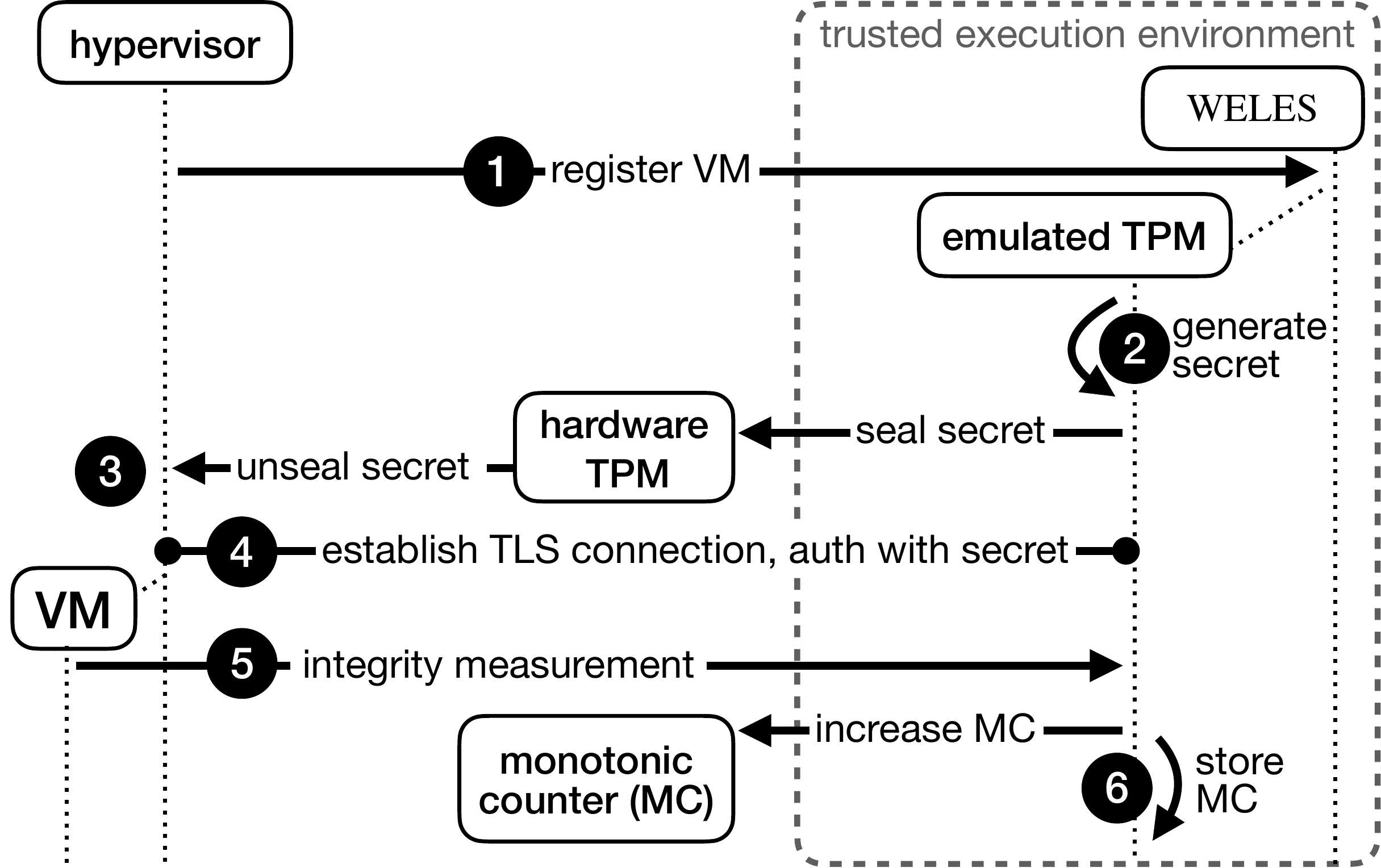}
\vspace{\captionvspacesize}
\caption{
    Attachment of an emulated TPM to a VM.
    \sys emulates \glspl{tpm} inside the \gls{tee}.
    Each emulated \gls{tpm} is accessible via a \gls{tls} connection.
    To prevent the \gls{mitm} attack, \sys authenticates the connecting hypervisor by sharing with him a secret via a hardware \gls{tpm}. 
    To mitigate the rollback attack, the emulated \gls{tpm} increments the monotonic counter value on each non-idempotent command.
}
\label{fig:vmstart}
\end{figure}
\autoref{fig:vmstart} shows the procedure of attaching an emulated TPM to a VM. Before the hypervisor spawns a VM, it commands \sys to emulate a new software-based TPM (\raisebox{-1pt}{\ding{202}}). \sys creates a new emulated TPM, generates a secret, and seals the secret with the hardware TPM (\raisebox{-1pt}{\ding{203}}). \sys returns the TCP port and the sealed secret to the hypervisor. The hypervisor unseals the secret from the hardware TPM (\raisebox{-1pt}{\ding{204}}) and establishes a TLS connection to the emulated TPM authenticating itself with the secret (\raisebox{-1pt}{\ding{205}}). At this point, the hypervisor spawns a \gls{vm}. The \gls{vm} boots up, the firmware and \gls{ima} send integrity measurements to the emulated \gls{tpm} (\raisebox{-1pt}{\ding{206}}). To protect against the rollback attack, each integrity measurement causes the emulated TPM to increment the hardware-based \gls{mc} and store the current \gls{mc} value inside the emulated TPM memory (\raisebox{-1pt}{\ding{207}}). To prevent the attachment of multiple VMs to the same emulated TPM, \sys permits a single client connection and does not permit reconnections. An attack in which an adversary redirects the hypervisor to a fake emulated TPM exporting a false secret is detected when establishing trust with the VM (\S\ref{sec:establish_trust}).

\subsection{Establish Trust}
\label{sec:establish_trust}
The tenant establishes trust with the \gls{vm} in three steps. First, he verifies that \sys executes inside the \gls{tee} and runs on genuine hardware (a CPU providing the \gls{tee} functionality). He then extends trust to the hypervisor and VM by leveraging \sys to verify and enforce the host and guest OSes' runtime integrity. Finally, he connects to the \gls{vm}, ensuring it is the \gls{vm} provisioned and controlled by \sys.

Since the \sys design does not restrict tenants to possess any vendor-specific hardware and the existing \gls{tee} attestation protocols are not standardized, we propose to add an extra level of indirection. Following the existing solutions~\cite{palaemon_2020}, we rely on a trusted \gls{ca} that performs the \gls{tee}-specific attestation before signing an X.509 certificate confirming the \sys's integrity and the underlying hardware genuineness. The tenant establishes trust with \sys during the \gls{tls}-handshake, verifying that the presented certificate was issued to \sys by the \gls{ca}.

Although the tenant remotely ensures that \sys is trusted, he has no guarantees that he connects to his VM controlled by \sys because the adversary can spoof the network~\cite{zheng_application-based_2009} to redirect the tenant's connection to an arbitrary \gls{vm}. To mitigate the threat, \sys generates a secret and shares it with the tenant and the VM. When the tenant establishes a connection, he uses the secret to authenticate the VM. Only the VM which integrity conforms to the policy has access to this secret.

\begin{figure}[tbp!]
    \centering
    \includegraphics[width=0.48\textwidth]{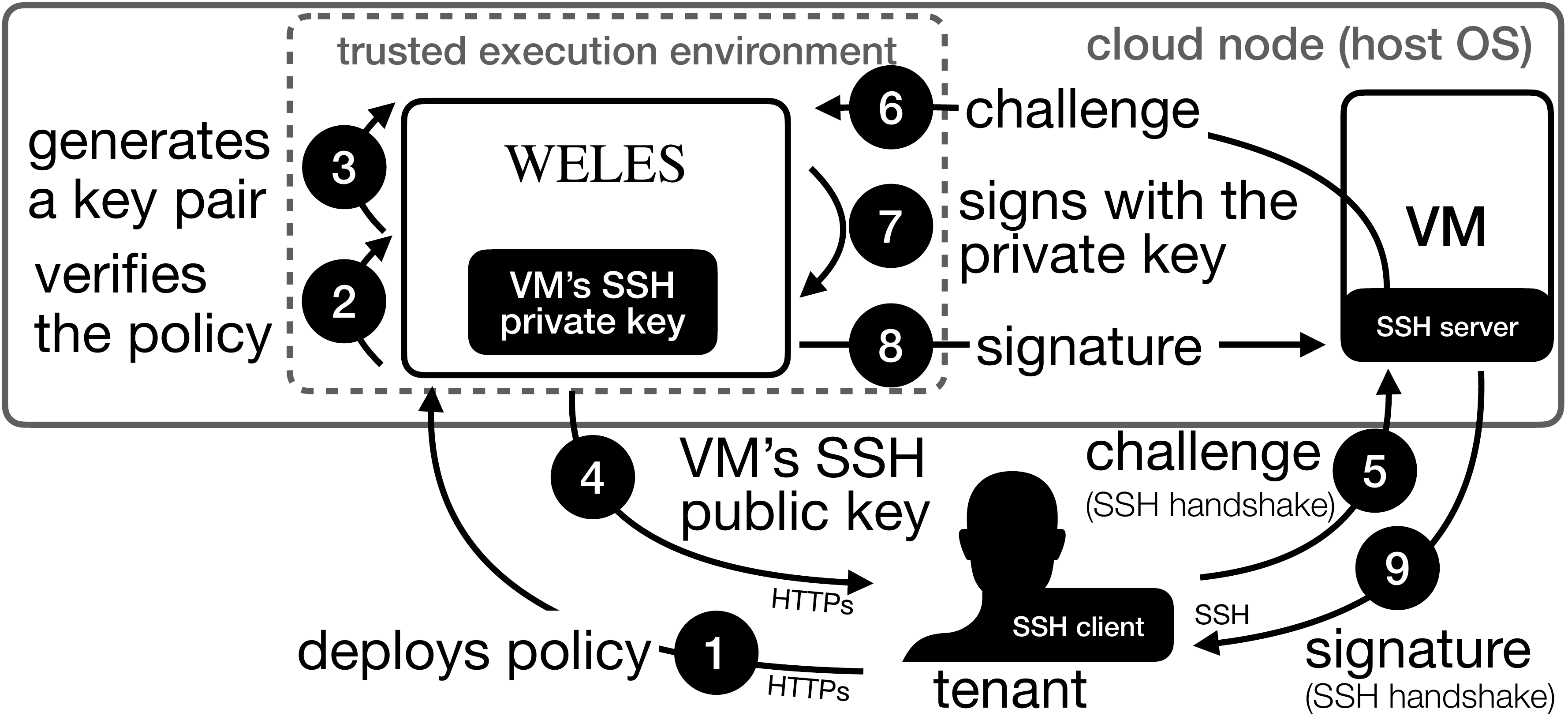}
    \vspace{\captionvspacesize}
    \caption{
        The high-level view of the attestation protocol. 
        \sys generates a SSH public/private key pair inside the \gls{tee}. 
        The tenant receives the public key as a result of the policy deployment.
        To mitigate the \gls{mitm} attacks, the tenant challenges the \gls{vm} to prove it has access to the private key.
        \sys signs the challenge on  behalf of the VM if and only if the platform integrity conforms with the policy.
    }
    \label{fig:ssh_dataflow}
\end{figure}
\autoref{fig:ssh_dataflow} shows a high-level view of the protocol. First, the tenant establishes a TLS connection with \sys to deploy the policy (\raisebox{-1pt}{\ding{202}}). \sys verifies the platform integrity against the policy (\raisebox{-1pt}{\ding{203}}), and once succeeded, it generates the \gls{ssh} key pair (\raisebox{-1pt}{\ding{204}}). The public key is returned to the tenant (\raisebox{-1pt}{\ding{205}}) while the private key remains inside the \gls{tee}. \sys enforces that only a guest OS which runtime integrity conforms to the policy can use the private key for signing. Second, the tenant initializes an \gls{ssh} connection to the VM, expecting the VM to prove the possession of the \gls{ssh} private key. The \gls{ssh} client requests the \gls{ssh} server running inside the VM to sign a challenge (\raisebox{-1pt}{\ding{206}}). The \gls{ssh} server delegates the signing operation to \sys (\raisebox{-1pt}{\ding{207}}). \sys signs the challenge using the private key (\raisebox{-1pt}{\ding{208}}) if and only if the hypervisor's and VM's integrity match the policy. The \gls{ssh} private key never leaves \sys; only a signature is returned to the \gls{ssh} server (\raisebox{-1pt}{\ding{209}}). The \gls{ssh} client verifies the signature using the \gls{ssh} public key obtained by the tenant from \sys (\raisebox{-1pt}{\ding{210}}). 
The SSH server also authenticates the tenant, who proves his identity using his own private SSH key. The SSH server is configured to trust his SSH public key. The tenant established trust in the remote platform as soon as the SSH handshake succeeds.

\subsection{Policy Enforcement}
\sys policy enforcement mechanism guarantees that the VM runtime integrity conforms to the policy. 
At the host OS, \sys relies on the IMA integrity-enforcement~\cite{ima_appraisal} to prevent the host kernel from loading to the memory files that are not digitally signed. 
Specifically, each file in the filesystem has a digital signature stored inside its extended attribute.
IMA verifies the signature issued by the cloud provider before the kernel loads the file to the memory.
The certificate required for signature verification is loaded from initramfs (measured by the DRTM) to the kernel's keyring.
At the guest OS, IMA inside the guest kernel requires the \sys approval to load a file to the memory.
The emulated TPM, controlled by \sys, returns a failure when IMA tries to extend it with measurement not conforming to the policy.
The failure instructs IMA not to load the file.

\section{Implementation}
\label{sec:implementation}

\subsection{Technology Stack}
We decided to base the prototype implementation on the Linux kernel because it is an open-source project supporting a wide range of hardware and software technologies.
It is commonly used in the cloud and, as such, can demonstrate the practicality of the proposed design.
QEMU~\cite{bellard_qemu_2005} and \gls{kvm}~\cite{kivity_kvm:2007} permit to use it as a hypervisor.
We rely on Linux \gls{ima}~\cite{ima_design_2004} as an integrity enforcement and auditing mechanism built-in the Linux kernel.

We chose Alpine Linux because it is designed for security and simplicity in contrast to other Linux distributions. 
It consists of a minimum amount of software required to provide a fully-functional OS that permits keeping a \gls{tcb} low.
All userspace binaries are compiled as position-independent executables with stack-smashing protection and relocation read-only memory corruption mitigation techniques. 
Those techniques help mitigate the consequences of, for example, buffer overflow attacks that might lead to privilege escalation or arbitrary code execution. 
To restrict the host from accessing guest memory and state, we follow existing security-oriented commercial solutions~\cite{ibm_secure_execution} that disable certain hypervisor features, such as hypervisor-initiated memory dump, huge memory pages on the host, memory swapping, memory ballooning through a virtio-balloon device, and crash dumps. 
For production implementations, we propose to rely on microkernels like formally proved seL4~\cite{klein_sel4:_2009}.

We rely on \gls{sgx} as the \gls{tee} technology. 
The \gls{sgx} remote attestation~\cite{johnson2016intel} allows us to verify if the application executes inside an enclave on a genuine Intel CPU. 
Other TEEs might be supported (\S\ref{sec:discussion}). 
We implemented \sys in Rust~\cite{matsakis_rust_2014}, which preserves memory-safety and type-safety. 
To run \sys inside an \gls{sgx} enclave, we use the SCONE framework~\cite{arnautov_scone:_2016} and its Rust cross-compiler. 
We also exploit the \gls{txt}~\cite{intel_txt_whitepaper} as a \gls{drtm} technology because it is widely available on Intel CPUs. 
We use the open-source software tboot~\cite{tboot} as a pre-kernel bootloader that establishes the \gls{drtm} with \gls{txt} to provide the measured boot of the Linux kernel. 

\begin{figure}[tbp!]
\centering
\includegraphics[width=0.48\textwidth]{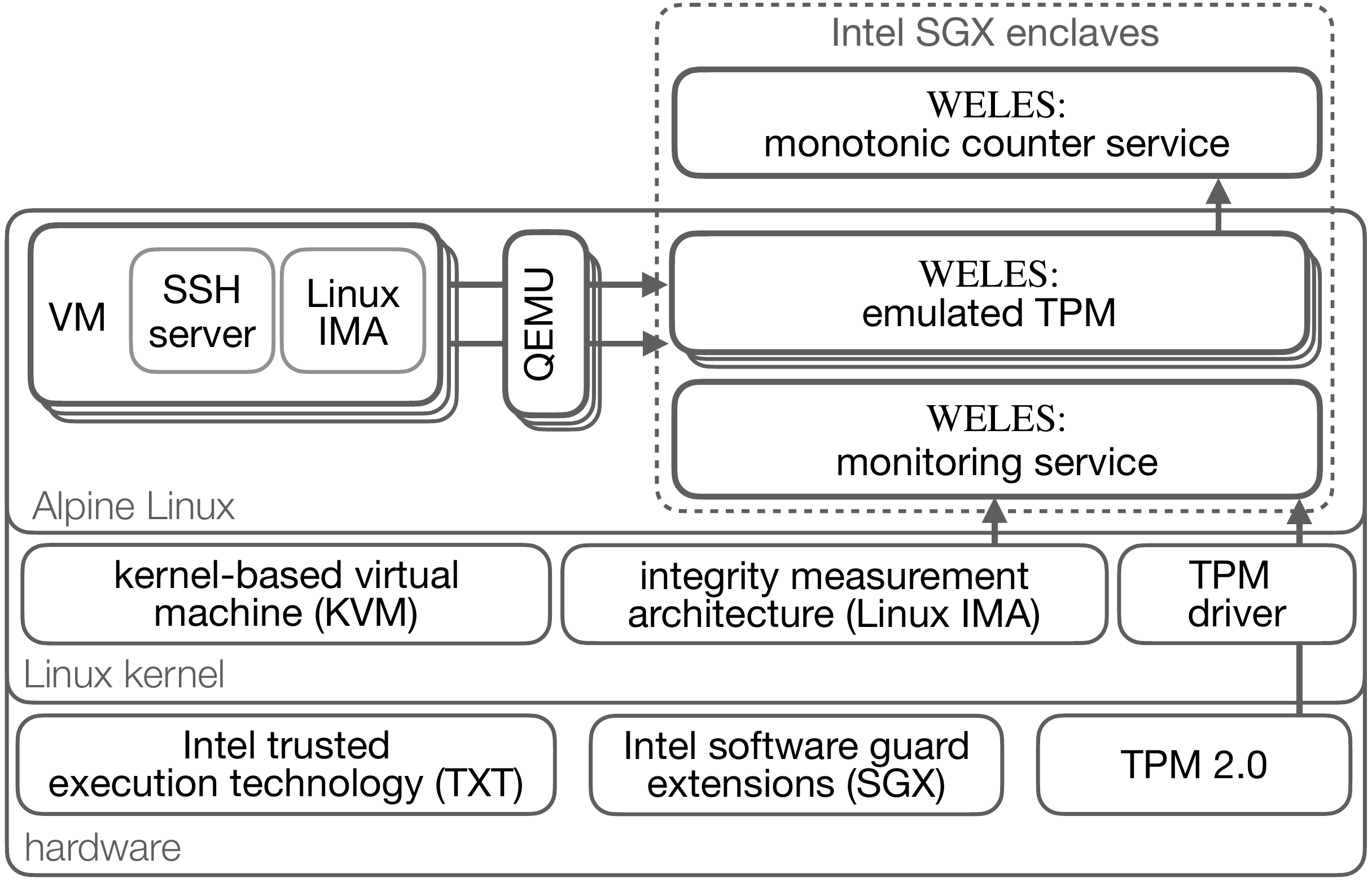}
\vspace{\captionvspacesize}
\caption{
    The overview of the \sys prototype implementation.
}
\label{fig:prototype}
\end{figure}

\subsection{Prototype Architecture}
The \sys prototype architecture consists of three components executing in \gls{sgx} enclaves: the monitoring service, the emulated TPM, and the monotonic counter service.

The monitoring service is the component that leverages Linux IMA and the hardware TPM to collect integrity measurements of the host OS.
There is only one monitoring service running per host OS.
It is available on a well-known port on which it exposes a TLS-protected REST API used by tenants to deploy the policy.
We based this part of the implementation on~\cite{tbe_2019} that provides a mechanism to detect the TPM locality.
The monitoring service spawns emulated TPMs and intermediates in the secret exchange between QEMU and the emulated TPM.
Specifically, it generates and seals to the hardware TPM the secret required to authenticate the QEMU process, and passes this secret to an emulated TPM. 

The emulated TPM is a software-based TPM emulator based on the libtpms library~\cite{libtpms}. It exposes a TLS-based API allowing QEMU to connect. The connection is authenticated using the secret generated inside an \gls{sgx} enclave and known only to processes that gained access to the hardware TPM. We extracted the emulated TPM into a separate component because of the libtpms implementation, which requires running each emulator in a separate process.

The \gls{mcs} provides access to a hardware \acrfull{mc}. 
Emulated TPMs use it to protect against rollback attacks. We designed the \gls{mcs} as a separate module because we anticipate that due to hardware \gls{mc} limitations (\ie, high latency, the limited number of memory overwrites~\cite{strackx_ariadne:_2016}), a distributed version, \eg, ROTE \cite{matetic_rote:2017}, might be required. Notably, the \gls{mcs} might also be deployed locally using \gls{sgx} \gls{mc}~\cite{intel_sgx_mc_whitepaper} accessible on the same platform where the monitoring service and emulated TPM run. 

\subsection{Monotonic Counter Service}
We implemented a \acrfull{mcs} as a service executed inside the \gls{sgx} enclave. 
It leverages TPM 2.0 high-endurance indices~\cite{tpm_2_0_spec_architecture} to provide the \gls{mc} functionality.
The \gls{mcs} relies on the TPM attestation to establish trust with the TPM chip offering hardware \gls{mc}, and on the encrypted and authenticated communication channel (\cite{tpm_2_0_spec_architecture}~\S19.6.7) to protect the integrity and confidentiality of the communication with the TPM chip from the enclave.
The \gls{mcs} exposes a REST API over a TLS (\S\ref{sec:tls_attestation}), allowing other enclaves to increment and read hardware monotonic counters remotely.

The emulated TPM establishes trust with the \gls{mcs} via the TLS-based SGX attestation (\S\ref{sec:tls_attestation}) and maintains the \gls{tls} connection open until the emulated TPM is shutdown.
We implemented the emulated TPM to increase the \gls{mc} before executing any non-idempotent TPM command, \eg, extending PCRs, generating keys, writing to non-volatile memory.
The \gls{mc} value and the \gls{tls} credentials are persisted in the emulated TPM state, which is protected by the \gls{sgx} during runtime and at rest via sealing.
When the emulated TPM starts, it reads the \gls{mc} value from the \gls{mcs} and then checks the emulated TPM state freshness by verifying that its \gls{mc} value equals the value read from the \gls{mcs}.

\subsection{TLS-based \gls{sgx} Attestation}
\label{sec:tls_attestation}
We use the SCONE key management system (CAS)~\cite{scone_cas} to perform remote attestation of \sys components, verify SGX quotes using \gls{ias}~\cite{anati2013innovative}, generate TLS credentials, and distribute the credentials and the CAS CA certificate to each component during initialization. 
\sys components are configured to establish mutual authentication over TLS, where both peers present a certificate, signed by the same CAS CA, containing an enclave integrity measurement. 
Tenants do not perform the SGX remote attestation to verify the monitoring service identity and integrity. 
Instead, they verify the certificate exposed by a remote peer during the policy deployment when establishing a TLS connection to the monitoring service. 
The production implementation might use Intel SGX-RA~\cite{intel_sgxra_whitepaper} to achieve similar functionality without relying on an external key management system.

\subsection{VM Integrity Enforcement}
\label{sec:impl:ima}
The current Linux IMA implementation extends the integrity digest of the \imalog entry to all active TPM PCR banks. For example, when there are two active PCR banks (\eg, SHA-1 and SHA-256), both are extended with the same value. We did a minor modification of the Linux kernel. It permitted us to share with the emulated TPM not only the integrity digest but also the file's measurement and the file's signature. We modified the content of the \emph{PCR\_Extend} command sent by the Linux IMA in a way it uses the SHA-1 bank to transfer the integrity digest, the SHA-256 bank to transfer the file's measurement digest, and the SHA-512 bank to transfer the file's signature. In the emulated TPM, we intercept the PCR\_extend command to extract the file's measurement and the file's digest. We use obtained information to enforce the policy; if the file is not permitted to be executed, the emulated TPM process closes the TLS connection causing the QEMU process to shut down the VM.

\subsection{SSH Integration}
To enable a secure connection to the VM, we relied on the OpenSSH server.
It supports the PKCS\#11~\cite{pkcs11_spec} standard, which defines how to communicate with cryptographic devices, like \gls{tpm}, to perform cryptographic operations using the key without retrieving it from the \gls{tpm}.

We configured an OpenSSH server running inside the guest OS to use an SSH key stored inside the emulated TPM running on the host OS.
The VM's SSH private key is generated and stored inside the SGX enclave, and it never leaves it.
The SSH server, via PCRS\#11, uses it for the TLS connection only when \sys authorizes access to it.
The tenant uses his own SSH private key, which is not managed by \sys.
\section{Evaluation}
\label{sec:evaluation}
\newcommand{\ubuntuver}{Ubuntu 18.04\xspace}
\newcommand{\alpinever}{Alpine 3.10\xspace}
\newcommand{\osver}{Alpine 3.10\xspace}

\newcommand{\kernelver}{Linux kernel 4.19\xspace}
\newcommand{\machinetype}{Dell PowerEdge R330\xspace}
\newcommand{\machinecpu}{Intel Xeon E3-1270 v5 CPU\xspace}
\newcommand{\tpmchipversion}{Infineon 9665 TPM 2.0\xspace}
\newcommand{\microcodever}{0x5e\xspace}

We present the evaluation of \sys in three parts. In the first part, we measure the performance of internal \sys components and related technologies. In the second part, we check if \sys is practical to protect popular applications, \ie, nginx, memcached. Finally, we estimate the \gls{tcb} of the prototype used to run the experiments mentioned above.

\myparagraph{Testbed.}
Experiments execute on a rack-based cluster of \machinetype servers equipped with an \machinecpu, 64\,GiB of RAM, \tpmchipversion \glspl{dtpm}. Experiments that use an \gls{itpm} run on Intel NUC7i7BNH machine, which has the \gls{ptt}~\cite{intel_ptt_whitepaper_2014} running on Intel ME 11.8.50.3425 powered by Intel Core i7-7567U CPU and 8\,GiB of RAM.
All machines have a 10\,Gb Ethernet \gls{nic} connected to a 20\,Gb/s switched network. 
The \gls{sgx}, \gls{txt}, \gls{tpm} 2.0, \gls{vtd} \cite{intel_vtd}, and \gls{sriov} \cite{intel_sriov} technologies are turned on in the UEFI/BIOS configuration. The hyper-threading is switched off. The \gls{epc} is configured to reserve 128\,MiB of \gls{ram}. 

On host and guest OSes, we run \osver with \kernelver. We modified the guest OS kernel according to the description in \S\ref{sec:impl:ima}. We adjusted \gls{qemu} 3.1.0 to support TLS-based communication with the emulated TPM as described in \S\ref{sec:impl:qemu}. CPUs are on the microcode patch level (\microcodever).

\subsection{Micro-benchmarks}
\label{micro}
\myparagraph{Are TPM monotonic counters practical to handle the rollback protection mechanism?}
Strackx and Piessens~\cite{strackx_ariadne:_2016} reported that the \gls{tpm} 1.2 memory gets corrupted after a maximum of 1.450M writes and has a limited increment rate (one increment per 5\,sec). We run an experiment to confirm or undermine the hypothesis that those limitations apply to the TPM 2.0 chip. We were continuously incrementing the monotonic counter in the \gls{dtpm} and the \gls{itpm} chips. The dTPM chip reached 85M increments, and it did not throttle its speed. The iTPM chip slowed down after 7.3M increments limiting the increment latency to 5\,sec. We did not observe any problem with the TPM memory. 

\begin{table}[b!]
\vspace{2mm}
\caption{
  The latency of main operations in the TPM-based \gls{mcs}. $\sigma$ states for standard deviation.
}
\vspace{\captiontablevspacesize}
\center
\begin{tabular}{lm{2.8cm}cc}
  \textbf{} & Read & Increase \\ \hline
discrete TPM & 42\,ms ($\sigma=2$\,ms) & 40\,ms ($\sigma=2$\,ms) \\
integrated TPM & 25\,ms ($\sigma=2$\,ms) & 32\,ms ($\sigma=1$\,ms) \\
\end{tabular}
\label{tab:eval:mc_latency}
\end{table}
\myparagraph{What is the cost of the rollback protection mechanism?}
Each non-idempotent TPM operation causes the emulated TPM to communicate with the MCS and might directly influence the \sys performance. We measured the latency of the TPM-based \gls{mcs} \emph{read} and \emph{increment} operations. In this experiment, the \gls{mcs} and the test client execute inside an \gls{sgx} enclave. Before the experiment, the test client running on the same machine establishes a \gls{tls} connection with the \gls{mcs}. The connection is maintained during the entire experiment to keep the communication overhead minimal. The evaluation consists of sending 5k requests and measuring the mean latency of the \gls{mcs} response.

\autoref{tab:eval:mc_latency} shows that the \gls{mcs} using \gls{itpm} performs from $1.25\times$ to $1.68\times$ faster than its version using \gls{dtpm}. The read operation on the \gls{itpm} is faster than the increment operation (25\,ms versus 32\,ms, respectively). Differently, on \gls{dtpm} both operations take a similar amount of time (about 40\,ms). 

\myparagraph{What is the cost of running the TPM emulator inside TEE and with the rollback protection mechanism? Is it slower than a hardware TPM used by the host OS?}
\begin{figure}[tbp!]
  \centering
  \includegraphics[width=0.5\textwidth]{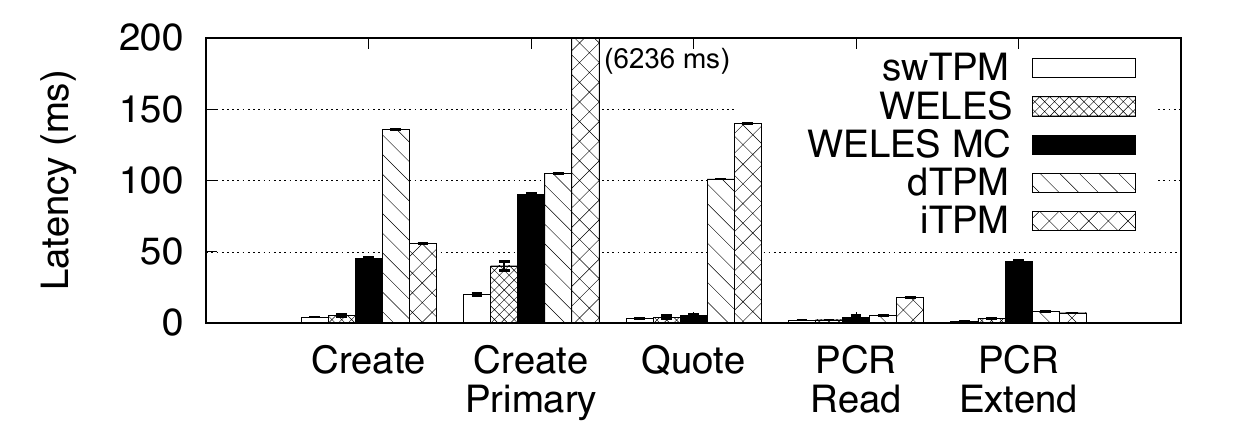}
  \vspace{-4mm}
  \caption{
    \Gls{tpm} operations latency depending on the TPM: the swTPM, the emulated TPM without rollback protection (\sys TPM), the emulated TPM with rollback protection (\sys TPM with MC), the \acrfull{dtpm}, and the \acrfull{itpm}.
  }
  \label{fig:tpm_operations}
\end{figure}

As a reference point to evaluate the emulated TPM's performance, we measured the latency of various \gls{tpm} commands executed in different implementations of TPMs. The TPM quotes were generated with the \gls{ecdsa} using the P-256 curve and SHA-256 bit key. PCRs were extended using the SHA-256 algorithm. \autoref{fig:tpm_operations} shows that except for the PCR extend operation, the SGX-based TPM with rollback protection is from $1.2\times$ to $69\times$ faster than hardware TPMs and up to $6\times$ slower than the unprotected software-based swTPM. Except for the \texttt{create primary} command, which derives a new key from the TPM seed, we did not observe performance degradation when running the TPM emulator inside an enclave. However, when running with the rollback protection, the TPM slows down the processing of non-idempotent commands (\eg, PCR\_Extend) due to the additional time required to increase the \gls{mc}.

\myparagraph{How much IMA impacts file opening times?}
\begin{figure}[tbp!]
  \centering
  \includegraphics[width=0.5\textwidth]{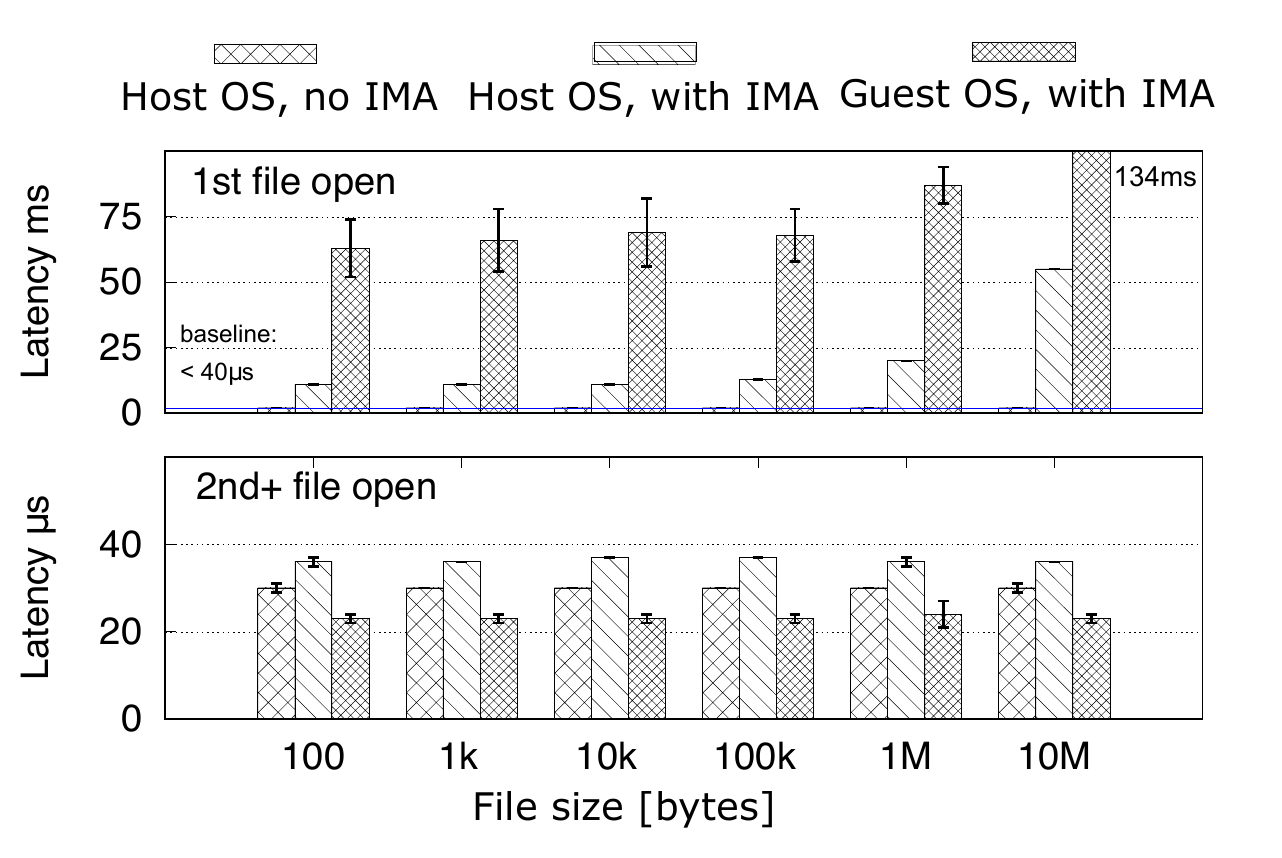}
  \vspace{-4mm}
  \caption{
	File opening times with and without Linux IMA.
  }
  \label{fig:load_times}
\end{figure}

Before the kernel executes the software, it verifies if executable, related configuration files, and required dynamic libraries can be loaded to the memory. The IMA calculates a cryptographic hash over each file (its entire content) and sends the hash to the TPM. We measure how much this process impacts the opening time of files depending on their size. \autoref{fig:load_times} shows that the IMA inside the guest OS incurs higher overhead than the IMA inside the host OS. 
It is primarily caused by i) the higher latency of the TPM extend command ($\sim43$\,ms) that is dominated by a slow network-based monotonic counter, ii) the IMA mechanism itself that has to calculate the cryptographic hash over the entire file even if only a small part of the file is actually read, and iii) the less efficient data storage used by the VM (virtualized storage, QCOW format). 
In both systems, the IMA takes less than 70\,ms when loading files smaller than 1\,MB (99\% of files in the deployed prototype are smaller than 1\,MB). Importantly, IMA measures the file only once unless it changes. \autoref{fig:load_times} shows that the next file reads take less than 40\,$\mu$s regardless of the file size.

\subsection{Macro-benchmarks}
\label{sec:eval:macro}
We run macro-benchmarks to measure performance degradation when protecting popular applications, \ie, the nginx web-server~\cite{nginx} and the memcached cache system~\cite{memcached}, with \sys. We compare the performance of four variants for each application running on the host \gls{os} (native), inside a SCONE-protected Docker container on the host \gls{os} (SCONE), inside a guest \gls{os} (VM), inside a \sys-protected guest \gls{os} with rollback protection turned on (\sys). Notably, \sys operates under a weaker threat model than SCONE. We compare them to show the tradeoff between the security and performance.

\myparagraph{How much does \sys influence the throughput of a web server, \eg, nginx?}
We configured nginx to run a single worker thread with turned off gzip compression and logging, according to available SCONE's benchmark settings. Then, we used wrk2~\cite{wrk2}, running on a different physical host, to simulate 16 clients (4 per physical core) concurrently fetching a pre-generated 10\,KiB binary uncompressed file for 45\,s.
\footnote{The limited network bandwidth dictated the file size. For larger size we saturated the NIC bandwidth.} 
We were increasing the frequency of the fetching until the response times started to degrade. Except for the reference measurement (native) run on the bare metal, nginx run inside a VM with access to all available cores and 4\,GB of memory. 

\begin{figure}[tbp!]
  \centering
  \includegraphics[width=0.5\textwidth]{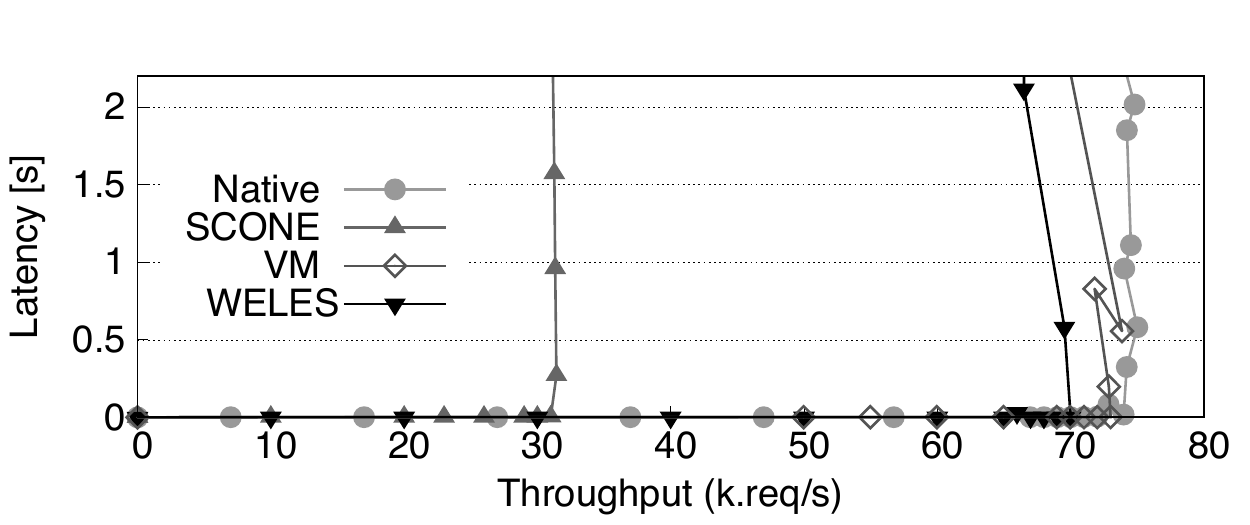}
  \vspace{-4mm}
  \caption{Throughput/latency of the \texttt{nginx}}  
  \label{fig:eval:nginx}
\end{figure}

\autoref{fig:eval:nginx} shows that \sys achieved $0.94\times$ of the native throughput, reaching 70k requests. The SCONE variant reached about 31k requests, which is $0.45\times$ of the \sys throughput. We observed low-performance degradation incurred by the virtualization (less than 2\%). The \sys overhead is caused mostly by the IMA. 

\myparagraph{Does \sys influence the throughput of systems that extensively use in-memory storage, \ie, memcached?}
\begin{figure}[bp!]
  \centering
  \includegraphics[width=0.5\textwidth]{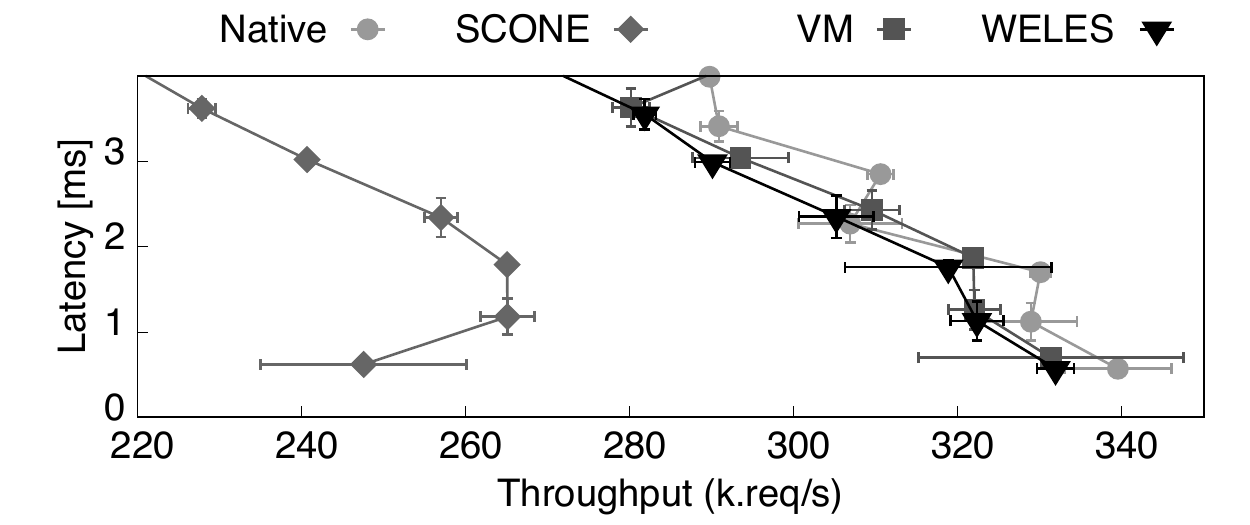}
  \vspace{-4mm}
  \caption{Throughput/latency of \texttt{memcached}}
  \label{fig:eval:memcached}
\end{figure}
We used memtier~\cite{memtier} to generate load by sending \texttt{GET}/\texttt{SET} requests (1:1 ratio) of 500\,bytes of random data to a memcached instance running on a different physical host. We calculated the memcached performance by computing the mean throughput achieved by the experiment before the throughput started to degrade (latency <2\,ms). Except for the reference measurement (native) run on the bare metal, memcached run in a VM with access to all available cores and 4\,GB of memory.
\autoref{fig:eval:memcached} shows \sys influence on the throughput-latency ratio of memcached. We observed low performance degradation when running memcached in a VM. \sys achieved $0.98\times$ of the native throughput. It is a result of IMA implementation, which measures the integrity of the memcached executables and configuration files during memcached launch, but it does not measure data directly written to the memory during runtime. \sys throughput was $1.23\times$ higher than the memcached run inside SCONE.

\myparagraph{How the measured boot increases the VM boot time?}
\newcommand{\YES}{{\ding{51}}}
\newcommand{\NO}{{\ding{55}}}
\begin{table}[b!]
    \vspace{2mm}
  \caption{
      The VM boot time duration when using different TPM implementations: no TPM, swTPM~\cite{swtpm_emulator}, emulated TPM inside an SGX enclave without rollback protection (\sys TPM), and emulated TPM inside an SGX enclave with rollback protection (\sys TPM with MC). $\sigma$ states for standard deviation.
  }
  \vspace{\captiontablevspacesize}
  \center
  \begin{tabular}{m{2.1cm}cccl}
    & MC & TPM & IMA & Boot time \\ \hline
  no TPM & \NO & \NO & \NO & 9.7 s ($\sigma=0.1$\,s) \\
  swTPM & \NO & \YES & \YES & 14.0 s ($\sigma=0.2$\,s) \\
  \sys & & & & \\
  \quad no MC & \NO & \YES & \YES & 14.1 s ($\sigma=0.3$\,s) \\
  \quad with MC & \YES & \YES & \YES & 50.8 s ($\sigma=0.4$\,s) \\
  \quad fast MC & \YES & \YES & \YES & 15.8 s (estimate) \\
  \hline
  \end{tabular}
  \label{tab:boot_time}
\end{table}
\autoref{tab:boot_time} shows how \sys impacts VM boot times. As a reference, we measure the boot time of a VM without any \gls{tpm} attached. Then, we run experiments in which a VM has access to different implementations of a software-based \glspl{tpm}. Except for the reference measurement, the Linux \gls{ima} is always turned on. Each VM has access to all available cores and 4\,GB of memory. As the guest OS, we run Ubuntu 18.10, a Linux distribution with a pre-installed tool (systemd-analyze) to calculate system boot times.
The measured boot increases the \gls{vm} load time. It is caused by the IMA module that measures files required to initialize the \gls{os}. We did not observe any difference in boot time between the setup with the swTPM~\cite{swtpm_emulator} and the \sys emulated TPM (\emph{\sys no MC}). However, when running the emulated TPM with the rollback protection (\emph{\sys with MC}), the VM boot time is $5.2\times$ and $3.6\times$ higher when compared to the reference and the swTPM setting, respectively. Alternative implementations of \gls{mcs}, such as ROTE \cite{matetic_rote:2017}, offer much faster MC increments (1--2\,ms) than the presented TPM-based prototype. We estimated that using \sys with a \emph{fast MC} would slow down VM boot time only by $1.13\times$.

\begin{figure}[tbp!]
  \centering
  \includegraphics[width=0.5\textwidth]{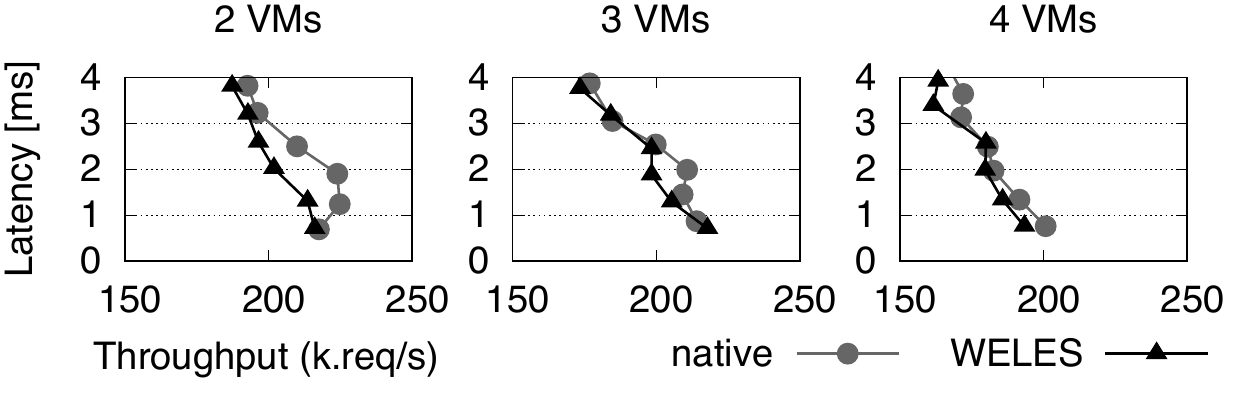}
  \vspace{-4mm}
  \caption{\sys scalability}
  \label{fig:eval:scalability:memcached}
\end{figure}
\myparagraph{Does \sys incur performance degradation when multiple VMs are spawned?} 
We run memcached concurrently in several VMs with and without \sys to examine \sys scalability. We then calculate the performance degradation between the variant with and without \sys. We do not compare the performance degradation between different number of VMs, because it depends on the limited amount of shared network bandwidth. 
Each VM run with one physical core and 1\,GB of RAM. \autoref{fig:eval:scalability:memcached} shows that when multiple VMs are concurrently running on the host OS, \sys achieves $0.96\times$--$0.97\times$ of the native throughput.

\begin{table}[tb!]
\caption{
    The \gls{tcb} (source code size in MB) of nginx and memcached when running in SCONE and \sys. All variants include the musl libc library.
}
\vspace{\captiontablevspacesize}
\center
\begin{tabular}{m{1.3cm}c|ccc}
  \textbf{} & \specialcell{+musl libc} & \specialcell{+SCONE} & \specialcell{+\sys} \\ 
\hline
Nginx & 7.3 & 12.3 ($1.6\times$) & 97.69 ($13.4\times$) \\
Memcached & 2.9 & 7.9 ($2.7\times$) & 93.29 ($32.2\times$) \\
\hline
\end{tabular}
\vspace{4mm}
\label{tab:eval:tcb_apps}
\end{table}

\subsection{Trusted Computing Base}
\label{sec:eval:tcb}
As part of this evaluation, we estimate the \gls{tcb} of software components by measuring the source code size (using \gls{sloc} as a metric) used to build the software stack.

\myparagraph{How much does the \sys increase the TCB?}
We estimate the software stack \gls{tcb} by calculating size of the source code metrics using CLOC~\cite{cloc_github}, lib.rs~\cite{lib_rs}, and Linux \texttt{stat} utility. 
For the Linux kernel and \gls{qemu}, we calculated code that is compiled with our custom configuration, \ie, Linux kernel with support for \gls{ima} and \gls{kvm}, and \gls{qemu} with support for gnutls, \gls{tpm} and \gls{kvm}.  

To estimate the \gls{tcb} of nginx and memcached executed with \sys, we added their source code sizes (including musl libc) to the source code size of the software stack required to run it (92.49\,MB). \autoref{tab:eval:tcb_apps} shows that the evaluated setup increases the \gls{tcb} $13\times$ and $32\times$ for nginx and memcached, respectively. When compared to SCONE, \sys prototype increases the \gls{tcb} from $8\times$ to $12\times$.
SCONE offers not only stronger security guarantees (confidentiality of the tenant's code and data) but also requires to trust a lower amount of software code.
However, this comes at the cost of performance degradation (\S\ref{sec:eval:macro}) and the necessity of reimplementing or at least recompiling a legacy application to link with the SCONE runtime.
\section{Discussion}
\label{sec:discussion}

\subsection{Alternative \glspl{tee}}
\label{sec:discussion_tees}
The \sys design (\S\ref{sec:overview}) requires a \gls{tee} that offers a remote attestation protocol and provides confidentiality and integrity guarantees of \sys components executing in the host OS.
Therefore, the \gls{sgx} used to build the \sys prototype (\S\ref{sec:implementation}) might be replaced with other \glspl{tee}.
In particular, \sys implementation might leverage Sanctum~\cite{sanctum_tee}, Keystone~\cite{keystone_tee}, Flicker~\cite{flicker2008}, or L4Re~\cite{reitz_l4re_tee} as an \gls{sgx} replacement.
\sys might also leverage ARM TrustZone~\cite{arm-trustzone} by running \sys components in the \emph{secure world} and exploiting the TPM attestation to prove its integrity. 

\subsection{Hardware-enforced VM Isolation}
Hardware CPU extensions, such as \gls{sev}~\cite{amd_sev_api}, \gls{mktme}~\cite{intel_mktme}, \gls{tdx}~\cite{intel_tdx_whitepaper}, are largely complementary to the \sys design. They might enrich \sys design by providing the confidentiality of the code and data against rogue operators with physical access to the machine, compromised hypervisor, or malicious co-tenants. They also consider untrusted hypervisor excluding it from the \sys \gls{tcb}.
On the other hand, \sys complements these technologies by offering means to verify and enforce the runtime integrity of guest OSes. Functionality easily available for bare-metal machines (via a hardware TPM) but not for virtual machines.

\subsection{Trusted Computing Base}
The prototype builds on top of software commonly used in the cloud, which has a large \gls{tcb} because it supports different processor architectures and hardware.
\sys might be combined with other \gls{tee} and hardware extensions, resulting in a lower \gls{tcb} and stronger security guarantees.
Specifically, \sys could be implemented on top of a microkernel architecture, such as formally verified seL4~\cite{klein_sel4:_2009, potts_mathematically_2014}, that provides stronger isolation between processes and a much lower code base (less than 10k \gls{sloc}~\cite{klein_sel4:_2009}), when compared to the Linux kernel. Comparing to the prototype, QEMU might be replaced with Firecracker~\cite{firecracker}, a virtual machine monitor written in a type-safe programming language that consists of 46k\,SLOC ($0.16\times$ of QEMU source code size) and is used in production by Amazon AWS cloud. 
The TCB of the prototype implementation might be reduced by removing superfluous code and dependencies. 
For example, most of the TPM emulator functionalities could be removed following the approach of $\mu$TPM~\cite{mccune_trustvisor:_2010}. 
\sys API could be built on top of the socket layer, allowing removal of HTTP dependencies that constitute 41\% of the prototype implementation code.

\subsection{Integrity Measurements Management}
The policy composed of digests is sensitive to software updates because newer software versions result in different measurement digests. Consequently, any software update of an integrity-enforced system would require a policy update, which is impractical. Instead, \sys supports dedicated \emph{update mirrors} serving updates containing digitally signed integrity measurements~\cite{tsr_2020, imasig_updates}. Other measurements defined in the policy can be obtained from the national software reference library~\cite{nist_hashes} or directly from the IMA-log read from a machine executed in a trusted environment, \eg, development environment running on tenant premises. The amount of runtime IMA measurements can be further reduced by taking into account processes interaction to exclude some mutable files from the measurement~\cite{ima_prima, infoflow_lsm}.
\section{Related work}
\label{sec:related}

\gls{vm} attestation is a long-standing research objective.
The existing approaches vary from \glspl{vm} monitoring systems focusing on system behavior verification~\cite{semantic_remote_attestation_vm, sadeghi_property-based_2004, poritz_property_2008}, intrusion detection systems~\cite{Joshi:2005:DPP:1095810.1095820, Sailer05shype:secure}, or verifying the integrity of the executing software~\cite{Garfinkel:2003:TVM:945445.945464, berger_scalable_2015}. 
\sys focuses on the VM runtime integrity attestation. 
Following Terra~\cite{Garfinkel:2003:TVM:945445.945464} architecture, \sys leverages \glspl{vm} to provide isolated execution environments constrained with different security requirements defined in a policy. 
Like Scalable Attestation~\cite{berger_scalable_2015}, \sys uses software-based \gls{tpm} to collect VM integrity measurements. 
\sys extends the software-based \gls{tpm} functionality by enforcing the policy and binding the attestation result with the \gls{vm} connection, as proposed by IVP~\cite{hutchison_verifying_2012}. 
Unlike the idea of linking the remote attestation quote to the \gls{tls} certificate~\cite{tpm_with_ca_vm_2006}, \sys relies on the \gls{tee} to restrict access to the private key based on the attestation result. 
Following TrustVisor~\cite{mccune_trustvisor:_2010}, \sys exposes trusted computing methods to legacy applications by providing them with dedicated TPM functionalities emulated inside the hypervisor. 
Unlike others, \sys addresses the TPM cuckoo attack at the VM level by combining integrity enforcement with key management and with the TEE-based remote attestation.
Alternative approaches to TPM virtualization exist~\cite{stumpf_enhancing_2008, ibm_tpm_commands}. However, the cuckoo attack remains the main problem. 
\sys enhances the vTPM design~\cite{perez_vtpm:_2006} mostly because of the simplicity; no need for hardware~\cite{stumpf_enhancing_2008} or the TPM specification \cite{ibm_tpm_commands} changes. 

Hardware solutions, such as \gls{sev}~\cite{amd_sev_api}, IBM PEF~\cite{hunt2021confidential}, \gls{tdx}~\cite{intel_tdx_whitepaper}, emerged to isolate VMs from the untrusted hypervisor and the cloud administrator.
However, they lack the VM runtime integrity attestation, a key feature provided by \sys.
\sys is complementary to them. 
Combining these technologies allows for better isolation of VM from the hypervisor and the administrator and for runtime integrity guarantees during the VM's runtime.

\section{Conclusion}
\label{sec:conclusion}
\glsresetall

This paper presented \sys, the \gls{vm} attestation protocol allowing for verification that security-sensitive applications execute in the VM composed and controlled by expected software in expected configuration. 
\sys provides transparent support for legacy applications, requires no changes in the \gls{vm} configuration, and permits tenants to remotely attest to the platform runtime integrity without possessing any vendor-specific hardware by binding the VM state to the SSH connection.
\sys complements hardware-based \gls{tee}, such as AMD SEV, IBM PEF, Intel TDX, by providing runtime integrity attestation of the guest OS. 
Finally, \sys incurs low performance overhead ($\le 6\%$) but has much larger \gls{tcb} compared to pure \gls{tee} approaches, such as \gls{sgx}.

\bibliographystyle{acm}
\interlinepenalty=100
\bibliography{bibliography}

\end{document}